\documentclass[apj]{emulateapj}

\begin{document}
\title{PTF10fqs: A Luminous Red Nova in the Spiral Galaxy Messier~99}
%\Author{Serene Saraswati \&\ Lax Laxmi} % Fun Names :-)

\author{Mansi M. Kasliwal, Shri R. Kulkarni,}
\affil{Cahill Center for Astrophysics, California Institute of Technology, Pasadena, CA, 91125, USA}

\author{Iair Arcavi,} 
\affil{Benoziyo Center for Astrophysics, Faculty of Physics, The Weizmann Institute of Science, Rehovot 76100, Israel}

\author{Robert M. Quimby, Eran O. Ofek,}
\affil{Cahill Center for Astrophysics, California Institute of Technology, Pasadena, CA, 91125, USA}

\author{Peter Nugent, Janet Jacobsen,}
\affil{Computational Cosmology Center, Lawrence Berkeley National Laboratory, 1 Cyclotron Road, Berkeley, CA 94720, USA}

\author{Avishay Gal-Yam, Yoav Green, Ofer Yaron,} 
\affil{Benoziyo Center for Astrophysics, Faculty of Physics, The Weizmann Institute of Science, Rehovot 76100, Israel}

\author{Derek B. Fox, Jacob L. Howell,}
\affil{Astronomy and Astrophysics, Eberly College of Science, The Pennsylvania State University, University Park, PA 16802, USA}

\author{S. Bradley Cenko, Io Kleiser, Joshua S. Bloom, Adam Miller, Weidong Li, Alexei V. Filippenko, Dan Starr,}
\affil{Department of Astronomy, University of California, Berkeley, CA 94720-3411, USA}

\author{Dovi Poznanski,}
\affil{Department of Astronomy, University of California, Berkeley, CA 94720-3411, USA}
\affil{Computational Cosmology Center, Lawrence Berkeley National Laboratory, 1 Cyclotron Road, Berkeley, CA 94720, USA}

\author{Nicholas M. Law,}
\affil{Dunlap Institute for Astronomy and Astrophysics, University of
Toronto, 50 St. George Street, Toronto M5S 3H4, Ontario, Canada}

\author{George Helou,}
\affil{Infrared Processing and Analysis Center, California Institute of Technology, Pasadena, CA 91125, USA}

\author{Dale A. Frail,}
\affil{National Radio Astronomy Observatory, Array Operations Center, Socorro, NM 87801, USA}

\author{James D. Neill, Karl Forster, D. Christopher Martin, Shriharsh P. Tendulkar,}
\affil{Cahill Center for Astrophysics, California Institute of
Technology, Pasadena, CA, 91125, USA}

\author{Neil Gehrels,}
\affil{NASA-Goddard Space Flight Center, Greenbelt, MD 20771, USA}

\author{Jamie Kennea,}
\affil{Department of Astronomy and Astrophysics, Pennsylvania State
University, State College, PA 16802, USA}

\author{Mark Sullivan,}
\affil{Department of Physics, Oxford University, Oxford, OX1 3RH, UK}

\author{Lars Bildsten,}
\affil{Kavli Institute of Theoretical Physics, University of California Santa Barbara, Santa Barbara, CA 93106, USA}
\affil{Department of Physics, University of California Santa Barbara, Santa Barbara, CA 93106, USA}

\author{Richard Dekany, Gustavo Rahmer, David Hale, Roger Smith, Jeff Zolkower, Viswa Velur, Richard Walters, John Henning, Kahnh Bui, Dan McKenna,}
\affil{Caltech Optical Observatories, California Institute of
Technology, Pasadena, CA 91125, USA}

%\centerline {and}

\author{Cullen Blake}
\affil{Department of Astrophysical Sciences, Princeton University, Princeton, NJ 08544, USA}

\bigskip
%\date{April 18 2010 -- \today} %% discovery to submission, 3 weeks! 

\begin{abstract}
The Palomar Transient Factory (PTF) is systematically charting 
the optical transient and variable sky. A primary science driver 
of PTF is building a complete inventory of transients
in the local Universe (distance less than 200 Mpc).  
%This is
%motivated by the needs of emerging areas of astrophysics (high
%energy cosmic rays,  neutrino and gravitational-wave astronomy) as
%well as curiosity of searching the large phase space between the
%luminosities and timescales of novae and supernovae.
Here, we report the discovery of PTF\,10fqs, a transient in the 
luminosity ``gap'' between novae and supernovae. 
Located on a spiral arm of Messier~99, PTF\,10fqs has a peak luminosity 
of $M_r = -12.3$, red color ($g-r=1.0$) and is slowly evolving (decayed by 1\,mag in 68\,days). It has a spectrum 
dominated by intermediate-width H$\alpha$ ($\approx$930\,km\,s$^{-1}$) and narrow calcium emission lines. 
The explosion signature (the light curve and spectra) is overall similar to
that of M85\,OT2006-1, SN\,2008S, and NGC\,300\,OT.  The origin of these
events is shrouded in mystery and controversy (and in some 
cases, in dust). PTF\,10fqs shows some evidence of a broad feature 
(around 8600\,\AA) that may suggest very large velocities 
($\approx$10,000\,km\,s$^{-1}$) in this explosion. 
%Another clue may be that PTF\,10fqs and SN\,2008S occurred in galaxies known to host multiple supernovae.
Ongoing surveys can be expected to find a few such events per year. 
Sensitive spectroscopy, infrared monitoring and statistics 
(e.g. disk versus bulge) will eventually make it possible 
for astronomers to unravel the nature of these mysterious explosions.
\end{abstract}

%{\it Facilities:} \facility{PO:1.2m ()}, \facility{Hale ()}, \facility{PO:1.5m ()}
%, \facility{HET ()}, \facility{Gemini:Gillett (), swift, galex, evla, hst, spitzer}

\section{Introduction}

Two reasons motivate us to search for transients in the local
Universe (distance $<$ 200\,Mpc). First, the emerging areas of
gravitational wave astronomy, high-energy cosmic rays, very high-energy 
photons, and neutrino astronomy are limited to this distance
horizon either due to physical effects (optical depth) or instrumental
sensitivity.
Thus, to effectively search for an electromagnetic
analog, understanding the full range of transient phenomena is
essential. For instance, the electromagnetic counterpart to the
gravitational wave signature of neutron star mergers 
is expected to be fainter and faster than that of supernovae
(e.g. \citealt{mmd+10}). 

Our second motivation is one of pure exploration.  The peak luminosity
of novae ranges between $-4$ and $-10$\,mag,\footnote{Unless
explicitly noted, quoted magnitudes are in the $R$ band} whereas supernovae
range between $-15$ and $-22$\,mag. 
%To date, the brightest
%classical nova was  $-$10 and the faintest thermonuclear supernova was
%$-$16 and the faintest core-collapse supernova was $-$14.5.  
The large gap between the cataclysmic novae and the catastrophic
supernovae has been noted by early observers. Theorists have
proposed several intriguing scenarios producing transients in this ``gap.''
(e.g. \citealt{bsw+07,mpq+09,skw+10,mtt+10})

The Palomar Transient 
Factory\footnote{\texttt{http://www.astro.caltech.edu/ptf.}} 
(PTF; see \citealt{gsv+08,lkd+09,rkl+09}) was designed to undertake
a systematic exploration of the transient sky in the optical
bands. One of the key projects of PTF is to build a 
complete inventory of transients in the
local Universe. PTF has a ``Dynamic'' cadence experiment which undertakes 
frequent observations of fields, optimized for inclusion of galaxies in the 
local Universe.  A description of the design sensitivity is given 
elsewhere \citep{kk09}.  Here, we report on the discovery of PTF\,10fqs,
a transient in this ``gap'' between novae and supernovae.

\section{Discovery}

\begin{figure}[htbp] 
   \centering
   \includegraphics[width=3in]{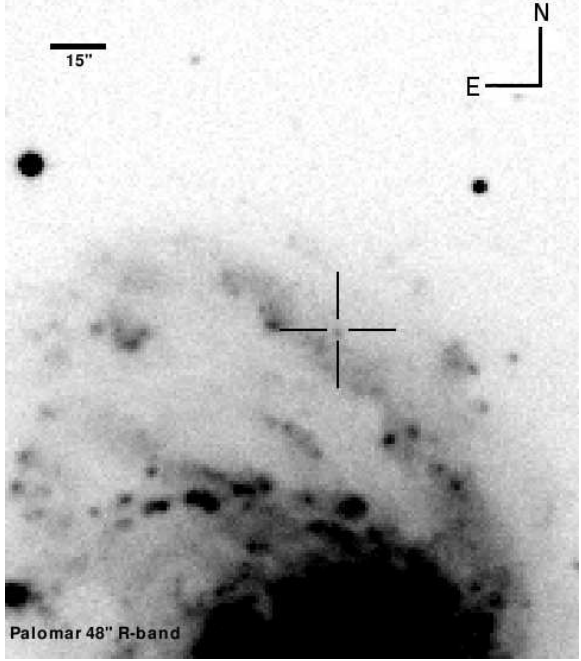} 
   \caption{\small The 
discovery image of PTF\,10fqs (obtained with the Palomar Oschin 48-inch telescope on 2010 Apr 16.393).
The transient is 
   marked by a cross and located at
   $\alpha$(J2000) = 12$^h$18$^m$50.16$^s$ and
   $\delta$(J2000) = $+14^\circ 26' 39.2''$. 
With respect to the host-galaxy nucleus, 
the transient is offset by 
8.1$^{\prime\prime}$\,E  and 99.9$^{\prime\prime}$\,N.}
\label{fig:DiscoveryImage}
\end{figure}

On 2010 April 16.393 (UT dates are used throughout this paper), 
the Palomar Transient Factory
discovered an optical transient toward Messier~99 (M99; see
Figure~\ref{fig:DiscoveryImage}).  Following the PTF discovery
naming sequence, this transient was dubbed PTF\,10fqs and reported
via an ATEL \citep{kk10}.

M99 (NGC~4254)\footnote{\texttt{http://seds.org/messier/m/m099.html}},
an Sc galaxy, is one of the brighter spiral members of the Virgo
cluster.  The recession velocity of the galaxy is about 
2400\,km\,s$^{-1}$.  Over the past fifty
years, three supernovae have been discovered in this galaxy: SN\,1967H
(Type II?, \citealt{f72}), SN\,1972Q (Type II; \citealt{bcr73}),{\footnote{Curiously, the reported
position of SN\,1972Q was only 3.6\arcsec\ from PTF\,10fqs. We did a
careful registration of the discovery image of SN\,1972Q \citep{bcr73} and 
PTF\,10fqs and find that the offset is actually 
11.0\arcsec\,E,0.8\arcsec\,S.}} and SN\,1986I (Type II; \citealt{pbc+89}).

At discovery, the brightness of PTF\,10fqs was $R = 20.0 \pm 0.2$\,mag.  There
are no previous detections in PTF data taken on and prior to April
10.  If located in M99, the absolute magnitude (for an assumed
distance of 17\,Mpc; \citealt{r02}) corresponds to $M_R = -11.1$
We concluded that the object could be (in decreasing order of probability)
a foreground variable star, a young supernova, or a transient
in the ``gap.'' These possibilities can be easily distinguished by
spectroscopic observations.

\section{Follow-Up Observations}

\subsection{Spectra}
\label{sec:Spectrum}

\begin{figure}[htbp] 
   \centering
   \includegraphics[height=3.5in,angle=90]{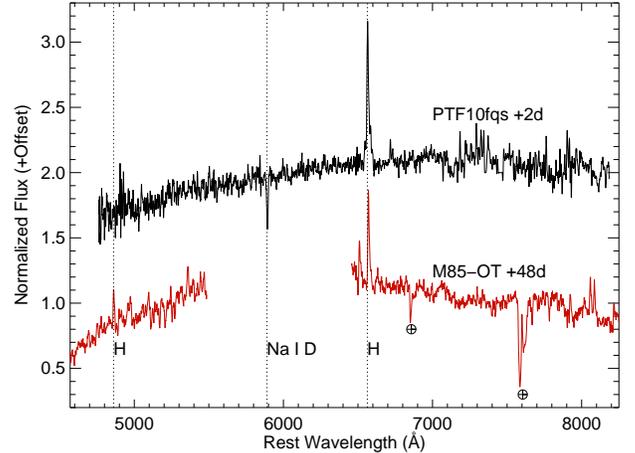} 
   \caption{\small Gemini GMOS
spectrum of PTF\,10fqs (black) taken two days 
after discovery. The wavelength
   coverage is continuous over the range 4600 to 8800\,\AA. 
   The most prominent emission feature is H$\alpha$. Plotted
   below for comparison, the spectrum of M85OT-2006-1 (red; \citealt{kor+07})} 
\label{fig:GMOSSpectrum}
\end{figure}

We triggered our Target-of-Opportunity (TOO) program on the
8-m Gemini-South telescope. On 2010 April 18.227, the Gemini 
Observatory staff observed PTF\,10fqs with the
Gemini Multi-Object Spectrograph (GMOS; \citealt{hja+04}).  The parameters for
the observations were: R400 grating, order-blocking filter GG455\_G039,
and a 1.0\arcsec\ slit.  Two 10-min integrations centered on
6700 and 6800\,\AA\ were obtained. The two observations allowed for
coverage of the gap between the chips. The package \texttt{gemini
gmos} working in the \texttt{iraf} framework was used to reduce the
data. The spectrum is shown in Figure~\ref{fig:GMOSSpectrum}.

The most prominent emission feature is an intermediate width (13\,\AA,
600\,km\,s$^{-1}$){\footnote{The velocity quoted here is corrected for instrumental 
resolution and is measured as the Gaussian Full Width Half Maximum 
(GFWHM) of the emission line.}} H$\alpha$ line consistent with the recession
velocity of the galaxy (2400\,km\,s$^{-1}$; see below). H$\beta$
was not detected.  
%(we confirm this with an inspection of the 2-d spectrum).  
From this spectrum alone, we concluded
that PTF\,10fqs is in M99 and the intermediate line width made it unlikely
to be a supernova.
PTF\,10fqs appeared to be a transient in the ``gap,'' and we initiated
extensive multi-band follow-up observations.

We continued to monitor the spectral evolution with the Marcario Low-Resolution Spectrograph (LRS; \citealt{hnm+98}) on the Hobby
Eberly Telescope{\footnote{Director's Discretionary Time, PI D. Fox.}}.
We used the G1 grating, with a 2\arcsec\ slit and a 
GG385 order-blocking filter, providing resolution 
$R  = \lambda/\Delta \lambda \approx 360$ over 4200--9200\,\AA. 
Data were reduced using the \texttt{onedspec} package in the
\texttt{iraf} environment, with cosmic-ray rejection via the \texttt{la\_cosmic} package \citep{v01}, and with
spectrophotometric corrections applied using standard-star observations (specifically,
BD332642). 

On May 15, we also obtained relatively higher resolution spectroscopic 
observations and relatively better blue coverage with the Low Resolution 
Imaging Spectrograph \citep{occ+95} on the Keck I telescope. 
First, we used the 831/8200 grating centered on 7905\,\AA\ to get higher 
resolution spectra of the Calcium 
lines. On the blue side, we used the 300/5000 grism to cover Ca H+K lines. 
For higher resolution covering the Balmer lines, we used the 600/7500 grating 
(centered on 7201\,\AA\,) in conjunction with the 600/4000 grism. 

\begin{figure*}[htbp] 
   \centering
   \includegraphics[height=6in,angle=90]{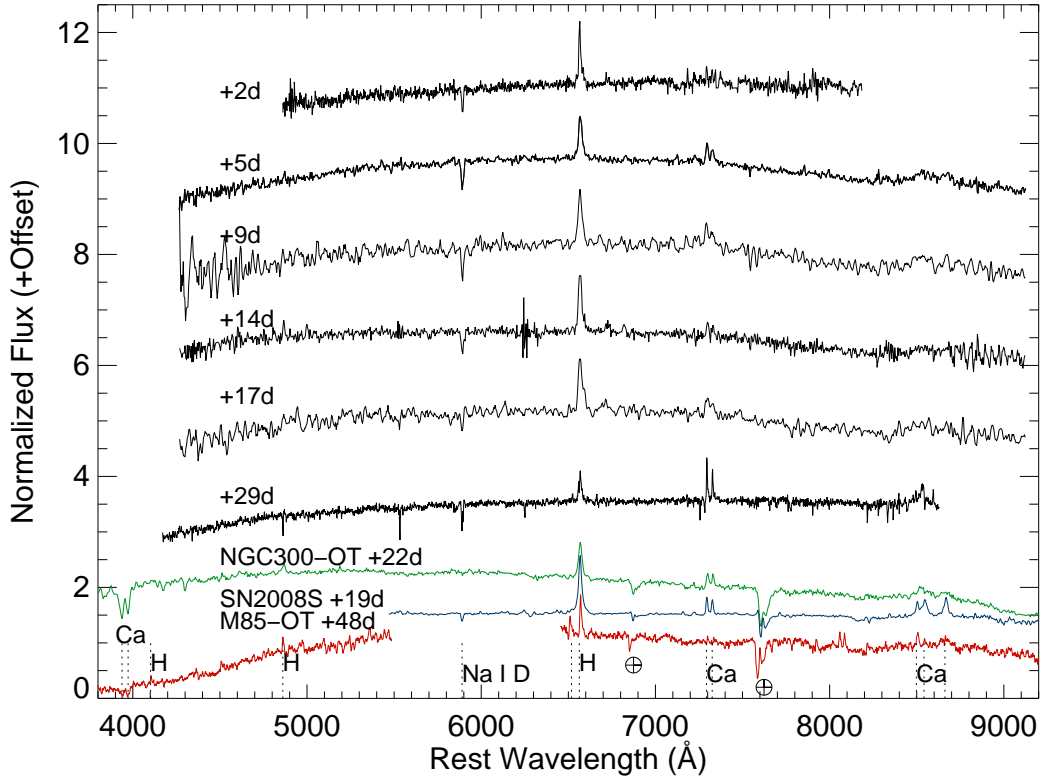} 
   \caption{Spectra of PTF\,10fqs at various epochs (phase in days is defined relative to discovery epoch). Also shown are spectra of NGC\,300-OT \citep{bbb+09}, M85OT2006-1 \citep{kor+07} and SN2008S \citep{bps+09}. The wavelength has been corrected 
for the recession velocity of each galaxy ($z=0.0024$ for M85, $z=0.008$ 
for M99, $z=0.00048$ for NGC\,300 and $z=0.00016$ for NGC\,6946).}
   \label{fig:Series}
\end{figure*}

The log of spectroscopic observations in given in Table~\ref{tab:spec}. The spectral evolution is shown in Figure~\ref{fig:Series}.

\begin{deluxetable}{lllll}[!hbt]
  \tabletypesize{\footnotesize}
  \tablecaption{Log of Spectroscopic Observations}
  \tablecolumns{5}
  \tablewidth{0pc}
 \tablehead{\colhead{Date (UT 2010)} & \colhead{MJD} & \colhead{Exposure} & \colhead{Facility} & \colhead{Grating/Grism} 
%\colhead{(UT 2010)} & \colhead{} & \colhead{}  & \colhead{}} \\
            }
  \startdata
Apr 18.23 & 55304.23 & 2$\times$ 600\,s & Gemini-S/GMOS & 400 \\
Apr 21.31 & 55307.31 & 2$\times$ 800\,s & HET/LRS & 360 \\
Apr 25.29 & 55311.29 & 2$\times$ 600\,s & HET/LRS & 360 \\
Apr 30.12 & 55316.12 & 2$\times$ 600\,s & HET/LRS & 360 \\
May 3.28  & 55319.28 & 2$\times$ 600\,s & HET/LRS & 360  \\
May 15.26 & 55331.26 & 3$\times$600\,s & Keck I/LRIS & 831 \\
May 15.26 & 55331.26 & 1$\times$2000\,s & Keck I/LRIS & 300 \\
May 15.31 & 55331.31 & 3$\times$650\,s & Keck I/LRIS & 600 
  \enddata
%  \tablecomments{ABC}
\label{tab:spec}
\end{deluxetable}

\subsection{Optical and Near-Infrared Imaging}

%{\it We need to wait a a couple of months to measure the photometric evolution.
%The coals in the fire are uBgriz photometry from Palomar 60-inch,
%JHK photometry from PAIRITEL, BVRI photometry from WISE}

\begin{figure*}[htbp] 
   \centering
   \includegraphics[height=4in,angle=0]{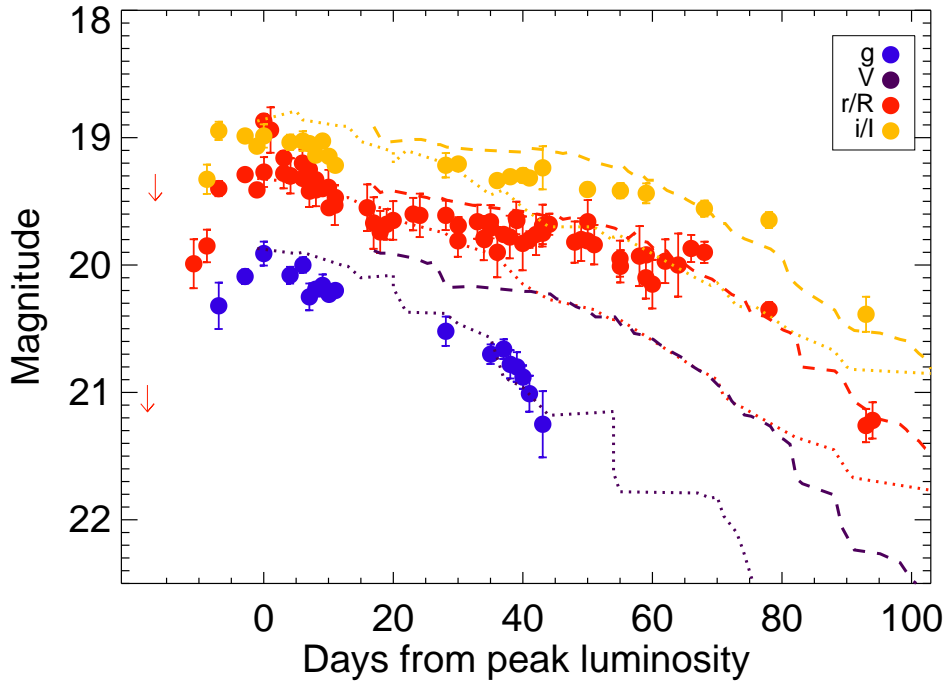} 
   \caption{Multi-band light curve of PTF\,10fqs obtained with the Palomar 48-inch (squares) and Palomar 60-inch (circles) telescopes. Upper limits are denoted by downward arrows. Note that the evolution is relatively faster in the $g$-band compared to $r$-band. Also shown for comparison are the $VRI$-band lightcurves of SN2008S (dotted; \citealt{bps+09}) and NGC300-OT (dashed; \citealt{bbb+09}). The
light curves are shifted vertically by a constant (+3\,mag for SN2008S and +5.2\,mag for NGC300-OT) such that their $R$-band light curves are at the same level as the $r$-band light curve of PTF\,10fqs.}
   \label{fig:lc}
\end{figure*}

Observations with the robotic Palomar 60-inch telescope \citep{cfm+06} on April 20.4 confirmed that
PTF\,10fqs was rising ($r = 19.4 \pm 0.1$ mag) and red ($g-r = 1.0$ mag).
We show the photometric evolution in $gri$-bands in Figure~\ref{fig:lc}
and Table~\ref{tab:grijhk}. On April 27.2, the light curve peaked at 
$r = 18.9 \pm 0.1$ mag corresponding to  $M_r = -12.3$ (correcting for 
foreground Galactic extinction  of E(B-V)=0.039; \citealt{sfd98}).  Aperture 
photometry was done after image 
subtraction using a custom modification of the CPM algorithm, 
{\it mkdifflc} \citep{gy+04}. Template images for subtraction
and reference magnitudes for zeropoint computation were 
taken from the Sloan Digital Sky Survey \citep{aaa+09}.

Near-infrared images were obtained with the Peters Automated Infrared
Imaging Telescope (PAIRITEL; \citealt{bsb+06}), and reduced by an
automated reduction pipeline. We lack sufficiently deep template
images, which are free of light from PTF\,10fqs, to perform reliable
image subtraction. Thus, we measure the flux from the source in a
small circular aperture, removing the sky with a nearby background 
region, and adopt a systematic error of 0.2 mag in the {\it J} and 
{\it H} bands and 0.3 mag in {\it $K_s$} band.  The values reported 
in Table~\ref{tab:grijhk} have been calibrated against the 2MASS system 
\citep{cwm03}.

\begin{longtable}{llll}[!hbt]
  \tabletypesize{\footnotesize}
  \tablecaption{Optical and Near-Infrared Light Curve}
  \tablecolumns{4}
  \tablewidth{0pc}
 \tablehead{\colhead{Date (MJD)} & \colhead{Filter} & \colhead{Mag} & \colhead{Facility} 
%\colhead{(UT 2010)} & \colhead{} & \colhead{}  & \colhead{}} \\
            }
%  \startdata
55295.2  &  $Mould$-$R$  &  $>$20.94  &  Palomar 48-in \cr
55296.5  &  $Mould$-$R$  &  $>$19.28  &  Palomar 48-in \cr
55302.4  &  $Mould$-$R$  &  19.99 $\pm$ 0.19  &  Palomar 48-in \cr
55313.2  &  $Mould$-$R$  &  19.27 $\pm$ 0.11  &  Palomar 48-in \cr
55316.3  &  $Mould$-$R$  &  19.28 $\pm$ 0.11  &  Palomar 48-in \cr
55317.3  &  $Mould$-$R$  &  19.30 $\pm$ 0.13  &  Palomar 48-in \cr
55319.2  &  $Mould$-$R$  &  19.20 $\pm$ 0.10  &  Palomar 48-in \cr
55320.2  &  $Mould$-$R$  &  19.42 $\pm$ 0.12  &  Palomar 48-in \cr
55321.3  &  $Mould$-$R$  &  19.41 $\pm$ 0.12  &  Palomar 48-in \cr
55323.2  &  $Mould$-$R$  &  19.39 $\pm$ 0.13  &  Palomar 48-in \cr
55324.2  &  $Mould$-$R$  &  19.53 $\pm$ 0.15  &  Palomar 48-in \cr
55329.2  &  $Mould$-$R$  &  19.55 $\pm$ 0.18  &  Palomar 48-in \cr
55330.2  &  $Mould$-$R$  &  19.67 $\pm$ 0.20  &  Palomar 48-in \cr
55331.2  &  $Mould$-$R$  &  19.74 $\pm$ 0.16  &  Palomar 48-in \cr
55332.2  &  $Mould$-$R$  &  19.68 $\pm$ 0.11  &  Palomar 48-in \cr
55333.2  &  $Mould$-$R$  &  19.65 $\pm$ 0.15  &  Palomar 48-in \cr
55336.3  &  $Mould$-$R$  &  19.60 $\pm$ 0.12  &  Palomar 48-in \cr
55337.3  &  $Mould$-$R$  &  19.61 $\pm$ 0.17  &  Palomar 48-in \cr
55343.2  &  $Mould$-$R$  &  19.81 $\pm$ 0.12  &  Palomar 48-in \cr
55346.2  &  $Mould$-$R$  &  19.66 $\pm$ 0.13  &  Palomar 48-in \cr
55347.2  &  $Mould$-$R$  &  19.79 $\pm$ 0.17  &  Palomar 48-in \cr
55348.2  &  $Mould$-$R$  &  19.66 $\pm$ 0.13  &  Palomar 48-in \cr
55349.3  &  $Mould$-$R$  &  19.90 $\pm$ 0.19  &  Palomar 48-in \cr
55351.2  &  $Mould$-$R$  &  19.78 $\pm$ 0.16  &  Palomar 48-in \cr
55352.2  &  $Mould$-$R$  &  19.63 $\pm$ 0.12  &  Palomar 48-in \cr
55353.2  &  $Mould$-$R$  &  19.83 $\pm$ 0.21  &  Palomar 48-in \cr
55355.2  &  $Mould$-$R$  &  19.76 $\pm$ 0.16  &  Palomar 48-in \cr
55356.2  &  $Mould$-$R$  &  19.69 $\pm$ 0.16  &  Palomar 48-in \cr
55361.2  &  $Mould$-$R$  &  19.82 $\pm$ 0.16  &  Palomar 48-in \cr
55362.2  &  $Mould$-$R$  &  19.80 $\pm$ 0.16  &  Palomar 48-in \cr
55363.2  &  $Mould$-$R$  &  19.66 $\pm$ 0.16  &  Palomar 48-in \cr
55364.2  &  $Mould$-$R$  &  19.84 $\pm$ 0.15  &  Palomar 48-in \cr
55368.2  &  $Mould$-$R$  &  19.95 $\pm$ 0.14  &  Palomar 48-in \cr
55371.2  &  $Mould$-$R$  &  19.93 $\pm$ 0.23  &  Palomar 48-in \cr
55372.2  &  $Mould$-$R$  &  20.10 $\pm$ 0.16  &  Palomar 48-in \cr
55373.2  &  $Mould$-$R$  &  20.15 $\pm$ 0.19  &  Palomar 48-in \cr
55375.2  &  $Mould$-$R$  &  19.97 $\pm$ 0.17  &  Palomar 48-in \cr
55377.2  &  $Mould$-$R$  &  20.00 $\pm$ 0.24  &  Palomar 48-in \cr
55379.2  &  $Mould$-$R$  &  19.87 $\pm$ 0.10  &  Palomar 48-in \cr
55304.4  &  $r$  &  19.85 $\pm$ 0.12  &  Palomar 60-in \cr
55306.3  &  $r$  &  19.40 $\pm$ 0.05  &  Palomar 60-in \cr
55310.3  &  $r$  &  19.29 $\pm$ 0.03  &  Palomar 60-in \cr
55312.1  &  $r$  &  19.41 $\pm$ 0.03  &  Palomar 60-in \cr
55313.2  &  $r$  &  18.87 $\pm$ 0.05  &  Palomar 60-in \cr
55314.2  &  $r$  &  18.94 $\pm$ 0.17  &  Palomar 60-in \cr
55316.3  &  $r$  &  19.16 $\pm$ 0.05  &  Palomar 60-in \cr
55317.3  &  $r$  &  19.30 $\pm$ 0.05  &  Palomar 60-in \cr
55319.2  &  $r$  &  19.32 $\pm$ 0.04  &  Palomar 60-in \cr
55320.2  &  $r$  &  19.25 $\pm$ 0.01  &  Palomar 60-in \cr
55321.2  &  $r$  &  19.33 $\pm$ 0.02  &  Palomar 60-in \cr
55322.3  &  $r$  &  19.40 $\pm$ 0.02  &  Palomar 60-in \cr
55323.3  &  $r$  &  19.55 $\pm$ 0.04  &  Palomar 60-in \cr
55324.3  &  $r$  &  19.47 $\pm$ 0.02  &  Palomar 60-in \cr
55341.3  &  $r$  &  19.61 $\pm$ 0.11  &  Palomar 60-in \cr
55343.2  &  $r$  &  19.69 $\pm$ 0.06  &  Palomar 60-in \cr
55347.3  &  $r$  &  19.80 $\pm$ 0.04  &  Palomar 60-in \cr
55348.2  &  $r$  &  19.71 $\pm$ 0.01  &  Palomar 60-in \cr
55350.2  &  $r$  &  19.76 $\pm$ 0.03  &  Palomar 60-in \cr
55352.3  &  $r$  &  19.65 $\pm$ 0.03  &  Palomar 60-in \cr
55354.2  &  $r$  &  19.80 $\pm$ 0.06  &  Palomar 60-in \cr
55356.3  &  $r$  &  19.75 $\pm$ 0.08  &  Palomar 60-in \cr
55357.3  &  $r$  &  19.68 $\pm$ 0.08  &  Palomar 60-in \cr
55363.2  &  $r$  &  19.81 $\pm$ 0.03  &  Palomar 60-in \cr
55368.3  &  $r$  &  20.01 $\pm$ 0.12  &  Palomar 60-in \cr
55372.2  &  $r$  &  19.92 $\pm$ 0.03  &  Palomar 60-in \cr
55381.2  &  $r$  &  19.90 $\pm$ 0.08  &  Palomar 60-in \cr
55391.2  &  $r$  &  20.35 $\pm$ 0.05  &  Palomar 60-in \cr
55406.2  &  $r$  &  21.26 $\pm$ 0.13  &  Palomar 60-in \cr
55407.2  &  $r$  &  21.22 $\pm$ 0.14  &  Palomar 60-in \cr
55306.3  &  $g$  &  20.32 $\pm$ 0.18  &  Palomar 60-in \cr
55310.3  &  $g$  &  20.09 $\pm$ 0.05  &  Palomar 60-in \cr
55313.2  &  $g$  &  19.91 $\pm$ 0.09  &  Palomar 60-in \cr
55317.3  &  $g$  &  20.08 $\pm$ 0.06  &  Palomar 60-in \cr
55319.2  &  $g$  &  20.00 $\pm$ 0.06  &  Palomar 60-in \cr
55320.2  &  $g$  &  20.25 $\pm$ 0.10  &  Palomar 60-in \cr
55321.2  &  $g$  &  20.19 $\pm$ 0.03  &  Palomar 60-in \cr
55322.3  &  $g$  &  20.16 $\pm$ 0.08  &  Palomar 60-in \cr
55323.3  &  $g$  &  20.23 $\pm$ 0.04  &  Palomar 60-in \cr
55324.3  &  $g$  &  20.20 $\pm$ 0.02  &  Palomar 60-in \cr
55341.3  &  $g$  &  20.52 $\pm$ 0.11  &  Palomar 60-in \cr
55348.2  &  $g$  &  20.70 $\pm$ 0.07  &  Palomar 60-in \cr
55350.2  &  $g$  &  20.66 $\pm$ 0.07  &  Palomar 60-in \cr
55351.3  &  $g$  &  20.78 $\pm$ 0.11  &  Palomar 60-in \cr
55352.3  &  $g$  &  20.80 $\pm$ 0.11  &  Palomar 60-in \cr
55353.2  &  $g$  &  20.88 $\pm$ 0.09  &  Palomar 60-in \cr
55354.2  &  $g$  &  21.01 $\pm$ 0.14  &  Palomar 60-in \cr
55356.3  &  $g$  &  21.25 $\pm$ 0.25  &  Palomar 60-in \cr
55304.4  &  $i$  &  19.32 $\pm$ 0.11  &  Palomar 60-in \cr
55306.3  &  $i$  &  18.94 $\pm$ 0.07  &  Palomar 60-in \cr
55310.3  &  $i$  &  18.98 $\pm$ 0.03  &  Palomar 60-in \cr
55312.2  &  $i$  &  19.06 $\pm$ 0.04  &  Palomar 60-in \cr
55313.2  &  $i$  &  18.98 $\pm$ 0.09  &  Palomar 60-in \cr
55317.3  &  $i$  &  19.03 $\pm$ 0.06  &  Palomar 60-in \cr
55319.2  &  $i$  &  19.02 $\pm$ 0.07  &  Palomar 60-in \cr
55320.2  &  $i$  &  19.04 $\pm$ 0.03  &  Palomar 60-in \cr
55321.2  &  $i$  &  19.13 $\pm$ 0.03  &  Palomar 60-in \cr
55322.3  &  $i$  &  19.02 $\pm$ 0.04  &  Palomar 60-in \cr
55323.3  &  $i$  &  19.14 $\pm$ 0.03  &  Palomar 60-in \cr
55324.2  &  $i$  &  19.21 $\pm$ 0.04  &  Palomar 60-in \cr
55341.3  &  $i$  &  19.21 $\pm$ 0.09  &  Palomar 60-in \cr
55343.2  &  $i$  &  19.20 $\pm$ 0.02  &  Palomar 60-in \cr
55349.2  &  $i$  &  19.33 $\pm$ 0.02  &  Palomar 60-in \cr
55351.3  &  $i$  &  19.30 $\pm$ 0.03  &  Palomar 60-in \cr
55353.2  &  $i$  &  19.29 $\pm$ 0.05  &  Palomar 60-in \cr
55354.2  &  $i$  &  19.31 $\pm$ 0.03  &  Palomar 60-in \cr
55356.3  &  $i$  &  19.23 $\pm$ 0.16  &  Palomar 60-in \cr
55363.2  &  $i$  &  19.40 $\pm$ 0.05  &  Palomar 60-in \cr
55368.3  &  $i$  &  19.41 $\pm$ 0.05  &  Palomar 60-in \cr
55372.2  &  $i$  &  19.43 $\pm$ 0.07  &  Palomar 60-in \cr
55381.2  &  $i$  &  19.55 $\pm$ 0.06  &  Palomar 60-in \cr
55391.2  &  $i$  &  19.64 $\pm$ 0.06  &  Palomar 60-in \cr
55406.2  &  $i$  &  20.38 $\pm$ 0.13  &  Palomar 60-in \cr
55307.2  &  $J$  &  18.14 $\pm$ 0.29  &  PAIRITEL \cr
55315.2  &  $J$  &  18.37 $\pm$ 0.39  &  PAIRITEL \cr
55317.2  &  $J$  &  17.89 $\pm$ 0.30  &  PAIRITEL \cr
55319.2  &  $J$  &  17.86 $\pm$ 0.26  &  PAIRITEL \cr
55321.2  &  $J$  &  17.94 $\pm$ 0.24  &  PAIRITEL \cr
55322.2  &  $J$  &  18.38 $\pm$ 0.25  &  PAIRITEL \cr
55324.2  &  $J$  &  17.88 $\pm$ 0.21  &  PAIRITEL \cr
55325.2  &  $J$  &  17.55 $\pm$ 0.32  &  PAIRITEL \cr
55327.2  &  $J$  &  17.86 $\pm$ 0.25  &  PAIRITEL \cr
55331.2  &  $J$  &  17.25 $\pm$ 0.18  &  PAIRITEL \cr
55333.2  &  $J$  &  17.82 $\pm$ 0.24  &  PAIRITEL \cr
55369.2  &  $J$  &  17.78 $\pm$ 0.31  &  PAIRITEL \cr
55307.2  &  $H$  &  17.35 $\pm$ 0.21  &  PAIRITEL \cr
55315.2  &  $H$  &  17.37 $\pm$ 0.27  &  PAIRITEL \cr
55317.2  &  $H$  &  17.14 $\pm$ 0.22  &  PAIRITEL \cr
55319.2  &  $H$  &  16.81 $\pm$ 0.27  &  PAIRITEL \cr
55321.2  &  $H$  &  17.75 $\pm$ 0.18  &  PAIRITEL \cr
55322.2  &  $H$  &  17.25 $\pm$ 0.16  &  PAIRITEL \cr
55324.2  &  $H$  &  17.22 $\pm$ 0.20  &  PAIRITEL \cr
55325.2  &  $H$  &  17.19 $\pm$ 0.30  &  PAIRITEL \cr
55327.2  &  $H$  &  17.02 $\pm$ 0.20  &  PAIRITEL \cr
55331.2  &  $H$  &  16.97 $\pm$ 0.32  &  PAIRITEL \cr
55333.2  &  $H$  &  17.07 $\pm$ 0.29  &  PAIRITEL \cr
55369.2  &  $H$  &  17.22 $\pm$ 0.22  &  PAIRITEL \cr
55307.2  &  $K$  &  16.17 $\pm$ 0.18  &  PAIRITEL \cr
55315.2  &  $K$  &  16.56 $\pm$ 0.31  &  PAIRITEL \cr
55317.2  &  $K$  &  16.84 $\pm$ 0.19  &  PAIRITEL \cr
55319.2  &  $K$  &  16.90 $\pm$ 0.25  &  PAIRITEL \cr
55321.2  &  $K$  &  16.84 $\pm$ 0.40  &  PAIRITEL \cr
55322.2  &  $K$  &  16.69 $\pm$ 0.21  &  PAIRITEL \cr
55324.2  &  $K$  &  16.29 $\pm$ 0.15  &  PAIRITEL \cr
55325.2  &  $K$  &  16.73 $\pm$ 0.18  &  PAIRITEL \cr
55327.2  &  $K$  &  16.65 $\pm$ 0.22  &  PAIRITEL \cr
55331.2  &  $K$  &  $>$15.80  &  PAIRITEL \cr
55333.2  &  $K$  &  $>$16.60  &  PAIRITEL \\
55369.2  &  $K$  &  $>$16.36  &  PAIRITEL 
%\enddata
\label{tab:grijhk}
\end{longtable}

\subsection{Radio Observations}
We observed PTF\,10fqs with the EVLA on April 20.19--20.26
at central frequencies of 4.96 GHz and 8.46 GHz. We added together two
adjacent 128 MHz subbands with full polarization to maximize continuum
sensitivity. Amplitude and bandpass calibration was achieved using
a single observation of J1331+3030, and phase calibration was carried
out every 10 min by switching between the target field and the
point source J1239+0730. The visibility data were calibrated and imaged
in the {\it AIPS} package following standard practice. 

A radio point
source was not detected at the position of the transient. After removing
extended emission from the host galaxy, the $3\sigma$ limits for a point
source are 93 $\mu$Jy and 63 $\mu$Jy at 4.96 GHz and 8.46 GHz, respectively.
At the distance of M99, this corresponds to $L_{\nu} <
2.1 \times 10^{25}$ erg s$^{-1}$ Hz$^{-1}$. Comparing with 
the compilation in \citet{cfn06}, this upper limit is 
at the level of the faintest Type II-P (SN\,2004dj; \citealt{bma+05}) and Type Ic (SN\,2002ap; \citealt{bkc02}) supernovae. As noted by \citet{bsc+09}, the nearby NGC300-OT was also not detected in the radio to deeper luminosity limits. 

\begin{figure}[htbp] 
   \centering
   \includegraphics[width=3.5in]{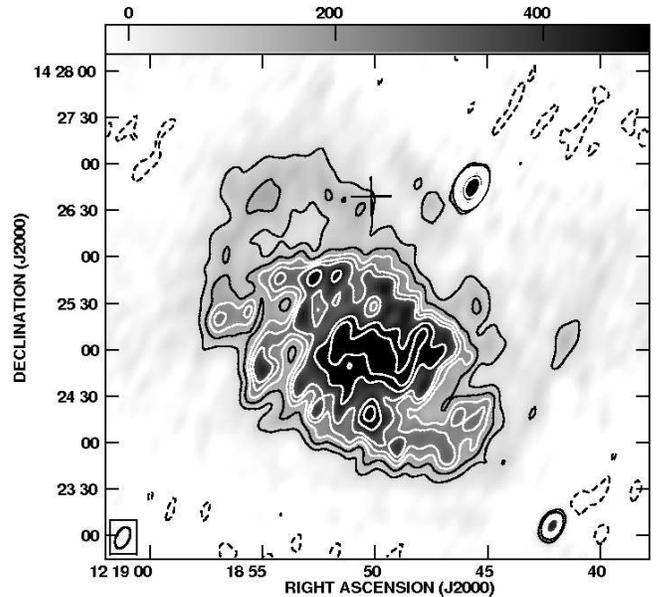} 
   \caption{Observation of PTF\,10fqs (denoted by a plus sign) with the 
EVLA at 4.96\,GHz, just four days after discovery. The gray-scale 
range is $-$40 to 1000\,$\mu$Jy per beam and the size of the 
synthesized beam is shown at the bottom-left corner.}
   \label{fig:ComparisonSpectrum}
\end{figure}

\subsection{Ultraviolet Observations}
We observed PTF\,10fqs with {\it GALEX} \citep{mfs+05} on two consecutive orbits
starting at 2010 April 24.387 (total exposure of 2846\,s).
All images were reduced and coadded using the standard {\it GALEX}
pipeline and calibration \citep{mcb+07}. 

\begin{figure}[htbp] 
   \centering
   \includegraphics[width=3.5in]{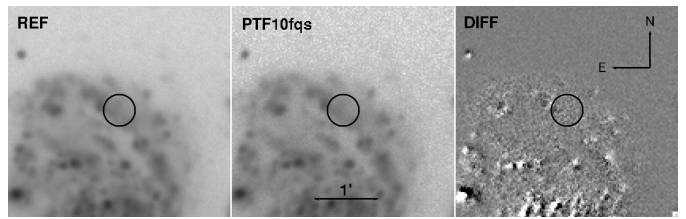} 
   \caption{Observation of PTF\,10fqs with {\it GALEX}. Reference data are 
taken from 22 images between 28 March 2005 and 2 April 2005 (left 
panel). Observations of PTF\,10fqs were taken on 24 April 2010 
(center panel). No source is detected in the difference image 
(right panel).}
   \label{fig:galex}
\end{figure}

To create a reference image, we coadded 22 images of M99 prior to 
2005 April 2 (total exposure of 18571\,s). Next, we subtracted the reference
image from observations of PTF\,10fqs (see Figure~\ref{fig:galex}). 
No source is detected. We find a 3$\sigma$ upper limit of NUV 22.7 AB
mag in an aperture consistent with a {\it GALEX} point source
($7.5\arcsec \times 7.5\arcsec$). 

To constrain the pre-explosion counterpart, we measured the limiting 
magnitude at the position of PTF\,10fqs in the coadded reference image. 
The faintest detected object consistent with being a point source 
within the galaxy had NUV = 20.1 AB mag. The 3$\sigma$ limit
based on measuring the sky root-mean square (rms) is NUV $>$ 21.8 AB mag.

\subsection{X-ray Observations}
We observed PTF10fqs with {\it Swift}/XRT on April 20.466 for 2507.3\,s 
and April 22.024 for 2623.5\,s. No source is detected
to a $3\sigma$ limiting count rate (assuming an $18''$ radius) of 
$4.6 \times 10^{-4}$ counts s$^{-1}$. Assuming a power-law model 
with a photon index of two, this corresponds to a flux limit of 
$1.6 \times 10^{-14}$ erg\,cm$^{-2}$\,s$^{-1}$.

\begin{deluxetable*}{lllllll}[!hbt]
  \tabletypesize{\footnotesize}
  \tablecaption{PTF\,10fqs Broadband Measurements}
  \tablecolumns{7}
  \tablewidth{0pc}
 \tablehead{\colhead{Date} & \colhead{MJD} & \colhead{Filter} &
             \colhead{Magnitude/Flux} & \colhead{$\nu$} & 
             \colhead{$\nu\,F_{\nu}$} & \colhead{Facility} \\
\colhead{(UT 2010)} & \colhead{ } & \colhead{ } & \colhead{ }  & \colhead{(Hz)} & \colhead{(erg cm$^{-2}$ s$^{-1}$)} & \colhead{ }
            }
  \startdata
%Apr 16.394 & $R$ (Vega) & 20.1 $\pm$ 0.2\,mag              &  4.848\,$\times$\,10$^{14}$ & 1.396\,$\times$\,10$^{-13}$ &  Palomar 48-in \\
Apr 20.23  & 55306.23 & 4.96 GHz & $<$93 $\mu$Jy               &  4.960\,$\times$\,10$^{9}$ & 4.613\,$\times$\,10$^{-18}$ &  EVLA \\
Apr 20.23  & 55306.23 & 8.46 GHz & $<$63 $\mu$Jy               &  8.460\,$\times$\,10$^{9}$ & 5.330\,$\times$\,10$^{-18}$ &  EVLA \\
%Apr 20.366 & $g$ (AB) & 21.1 $\pm$ 0.2\,mag                &  6.354\,$\times$\,10$^{14}$ & 8.260\,$\times$\,10$^{-14}$ &  Palomar 60-in \\
Apr 20.466 & 55306.466 & 0.3--10 keV & $<$4.6$\times$10$^{-4}$\,cps & 4.200\,$\times$\,10$^{17 }$ &2.864\,$\times$\,10$^{-15}$  &  {\it Swift}/XRT \\
%Apr 21.22 & $J$ (Vega) & 17.9 $\pm$ 0.2\,mag              &  2.418\,$\times$\,10$^{14}$& 2.636\,$\times$\,10$^{-13}$ &  PAIRITEL\\
%Apr 21.22 & $H$ (Vega) & 17.3 $\pm$ 0.2\,mag                 &  1.817\,$\times$\,10$^{14}$ & 2.144\,$\times$\,10$^{-13}$ &  PAIRITEL \\
%Apr 21.22 & $K_s$ (Vega) & 16.2 $\pm$ 0.3\,mag             &  1.368\,$\times$\,10$^{14}$ & 2.928\,$\times$\,10$^{-13}$ &  PAIRITEL\\
Apr 24.646 & 55310.646 & NUV (AB)  & $>$22.7\,mag                      &  1.295\,$\times$\,10$^{15}$ & 3.885\,$\times$\,10$^{-14}$ &  {\it GALEX} 
%May 3.221 & $R$ (Vega) & 19.4 $\pm$ 0.2\,mag                 &  4.848\,$\times$\,10$^{14}$ & 2.657\,$\times$\,10$^{-13}$ &  Palomar 48-in 
 \enddata
%  \tablecomments{ABC}
\label{tab:lc}
\end{deluxetable*}

\section{Archival Data}

\subsection{Hubble Space Telescope (HST)}
A query to the Hubble Legacy archive returned {\it HST} images of M99 in
the F606W (2001), F336W (2009), and F814W (2009) filters. 
We multidrizzled this data (PI Regan, Proposal ID 11966) and registered 
our Gemini/GMOS acquisition
image with the {\it HST/WFPC2} images. Unfortunately, PTF\,10fqs is just off 
the edge of the chip for the F606W filter image. 

\begin{figure*}[!htbp] 
   \centering
   \includegraphics[width=4in]{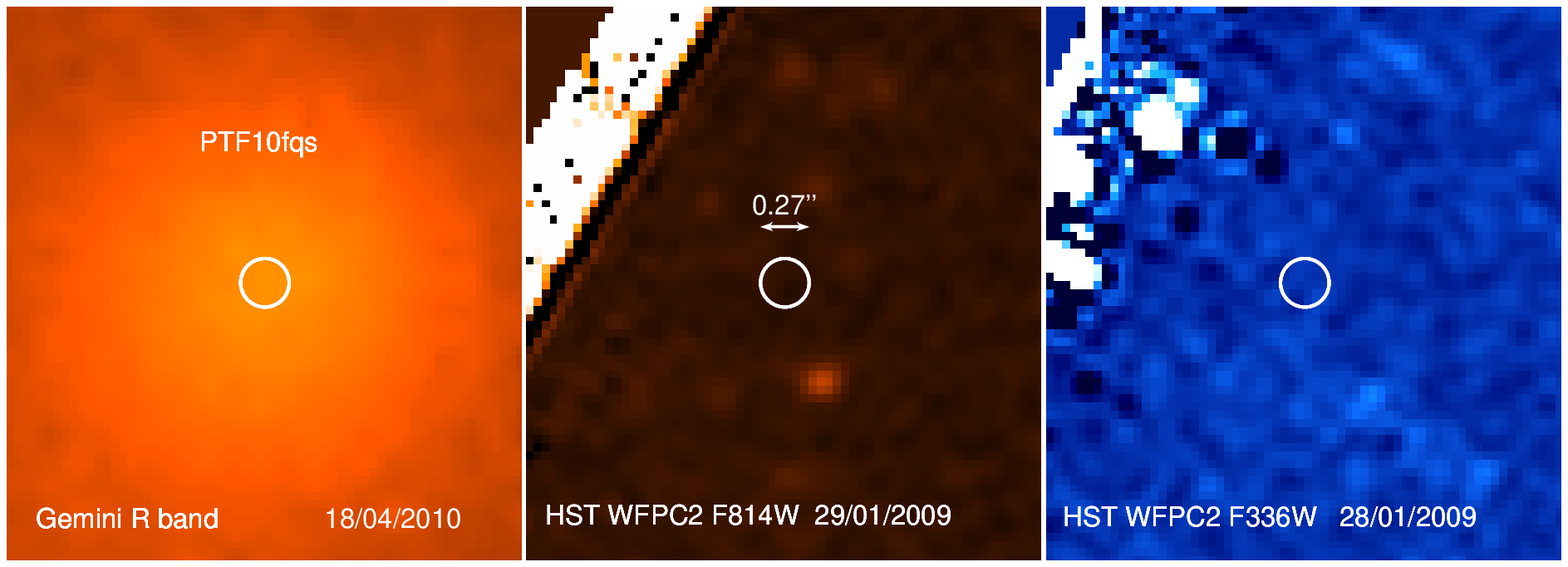} 
    \includegraphics[width=4in]{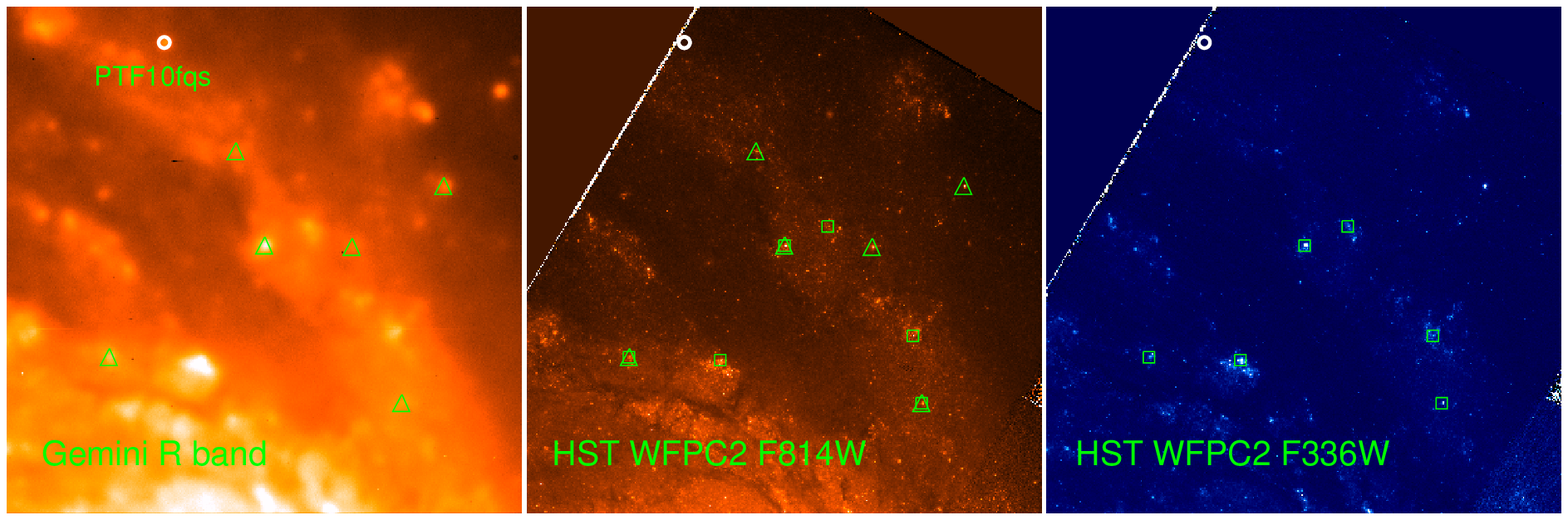} 
   \caption{HST/F814W and HST/F336W observations from 2009. 
{\it Top panel:} Zoomed-in view (2.8$\arcsec \times$ 2.6$\arcsec$) to show the absence of a pre-explosion 
counterpart. This rules out red supergiants fainter than $M_V = -3$\,mag and blue supergiants 
fainter than $M_V = -4.3$\,mag. {\it Bottom panel:} Zoomed-out view (81.2$\arcsec \times$ 82.1$\arcsec$) to 
show registration stars. Stars used to register the Gemini/R-band image with the HST/F814W image are denoted 
by triangles. Stars used to register the HST/F814W image with the HST/F336W are denoted by squares.}
   \label{fig:hst}
\end{figure*}

The total 1$\sigma$ registration error, added in quadrature, was 0.59 pixels.
There sources of error are as follows: centroiding error (0.17 in x, 0.30 in y), 
registration error between the Gemini image and HST/F814W image (0.19 in x, 0.44 in y)
and registration error between HST/F814W image and HST/F336W image (0.04 in x, 0.02 in y). 
Hence, in Figure~\ref{fig:hst} we plot a 5$\sigma$ radius of 3 pixels or 0.27$\arcsec$. 

No source is detected at the location of PTF\,10fqs. To estimate the limiting
magnitude, we ran sextractor and performed photometry following \citet{hbc+95}.
We find 3$\sigma$ limiting Vega magnitudes of $I > 26.9$ and $U > 26$ in the
1800\,s and 6600\,s exposures respectively.  

\subsection{Spitzer Space Telescope}

\begin{figure}[htbp] 
   \centering
   \includegraphics[width=3.5in]{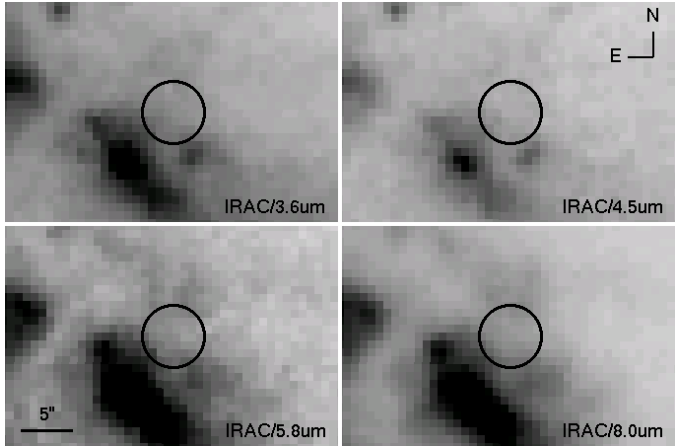} 
   \caption{Pre-explosion observations with {\it Spitzer}/IRAC. No source is found to be consistent with PTF\,10fqs.}
   \label{fig:Spitzer}
\end{figure}

M99 was part of the sample of the SIRTF Nearby Galaxies Survey (SINGS) galaxies \citep{kab+03}. 
This program undertook IRAC and MIPS imaging in 2004--2005. 
No point source is detected at the location of PTF\,10fqs (see Figure~\ref{fig:Spitzer}). 
We downloaded IRAC images from the final data release of SINGS  
and MIPS images from the standard {\it Spitzer} pipeline. 
Computed upper limits
(see Table~\ref{tab:progenitor}) assume
a 2-pixel aperture radius and sky-rms based on a $20 \times 20$ pixel box at the 
location. 

%The 3$\sigma$ limiting magnitude of these images constrains a progenitor to
%be fainter than $-$13.2, $-$14.2, $-$16.2 and $-$16.6 in 3.6\,$\mu$m, 4.5\,$\mu$m,  
%5.8\,$\mu$m and 8.0\,$\mu$m respectively. The conversion from Spitzer flux to
%Vega magnitude was done using zeropoints available on G. Wilson's webpage
%{\footnote{http://web.ipac.caltech.edu/staff/gillian/cal.html}}. 

\begin{deluxetable}{llll}[!hbt]
  \tabletypesize{\footnotesize}
  \tablecaption{Progenitor constraints for PTF\,10fqs}
  \tablecolumns{4}
  \tablewidth{0pc}
 \tablehead{\colhead{Date} & \colhead{Filter} &
             \colhead{Magnitude/Flux} & \colhead{Facility} 
%\colhead{(UT 2010)} & \colhead{} & \colhead{}  & \colhead{}} 
            }
  \startdata
2005  & NUV (AB) & $>$21.8\,mag & {\it GALEX} \\
2009  & F336W (Vega $U$) & $>$26\,mag & {\it HST}/WFPC2 \\
2009  & F814W (Vega $I$) & $>$26.9\,mag & {\it HST}/WFPC2 \\
2004  & 3.6\,$\mu$m & $<$5.3 $\mu$Jy & {\it Spitzer}/IRAC \\
2004  & 4.5\,$\mu$m & $<$3.5 $\mu$Jy & {\it Spitzer}/IRAC \\
2004  & 5.8\,$\mu$m & $<$51 $\mu$Jy & {\it Spitzer}/IRAC \\
2004  & 8.0\,$\mu$m & $<$344 $\mu$Jy & {\it Spitzer}/IRAC \\ 
2004  & 23.68\,$\mu$m & $<$240 $\mu$Jy & {\it Spitzer}/MIPS
%2004  & 71.42\,$\mu$m & $<$68 mJy & {\it Spitzer}/MIPS \\
%2004  & 155.9\,$\mu$m & $<$230 mJy & {\it Spitzer}/MIPS 
 \enddata
%  \tablecomments{ABC}
\label{tab:progenitor}
\end{deluxetable}

\subsection{Katzman Automatic Imaging Telescope}
\label{sec:kait}
The 0.76\,m Katzman Automatic Imaging Telescope
(KAIT{\footnote{http://astro.berkeley.edu/$\sim$bait/kait.html.}};
\citealt{lft+00,flt+01})
had extensively imaged M99 in the past decade --- 
113 images in the period 1999--2010. We stacked the
images in each season and find no point source at 
the location of PTF\,10fqs. Limiting magnitudes for each
season are summarized in Table~\ref{tab:kait}.

\begin{deluxetable}{llll}[!hbt]
  \tabletypesize{\footnotesize}
  \tablecaption{Historical Optical Observations}
  \tablecolumns{4}
  \tablewidth{0pc}
 \tablehead{\colhead{Date Range} & \colhead{Exposure} &
             \colhead{Limiting Mag} & \colhead{Facility}\\
\colhead{(UT)} & \colhead{(seconds)} & \colhead{($R$ band)} & \colhead{ }
} 
  \startdata
1998-12-27 -- 1999-06-01  &  680.0  &  $>$ 20.4 & KAIT \\
1999-11-26 -- 2000-06-07  &  567.0  &  $>$ 20.4 & KAIT \\
2001-04-11 -- 2001-06-07  &  192.0  &  $>$ 20.1 & KAIT \\
2002-01-14 -- 2002-06-08  &  486.0  &  $>$ 20.4 & KAIT \\
2003-01-15 -- 2003-06-04  &  318.0  &  $>$ 20.4 & KAIT \\
2004-01-29 -- 2004-06-16  &  392.0  &  $>$ 20.3 & KAIT \\
2004-12-25 -- 2005-06-01  &  110.0  &  $>$ 20.3 & KAIT \\
2006-01-12 -- 2006-05-18  &  665.7  &  $>$ 22.2 & DeepSky \\
2006-03-24 -- 2006-05-18  &  78.0   &  $>$ 20.4 & KAIT \\
2007-01-04 -- 2007-05-06  &  1749.9 &  $>$ 22.4 & DeepSky \\
2007-01-13 -- 2007-06-04  &  178.0  &  $>$ 20.4 & KAIT \\
2007-12-22 -- 2008-06-16  &  332.0  &  $>$ 20.4 & KAIT \\
2008-05-18 -- 2008-05-18  &  241.2  &  $>$ 20.7 & DeepSky \\
2009-03-28 -- 2009-04-27  &  64.0   &  $>$ 20.3 & KAIT \\
2010-02-11 -- 2010-03-22  &  32.0   &  $>$ 20.0 & KAIT
 \enddata
  \tablecomments{All images in a season were stacked.}
\label{tab:kait}
\end{deluxetable}

\subsection{DeepSky imaging}
\label{sec:deepsky}
DeepSky{\footnote{http://supernova.lbl.gov/$\sim$nugent/deepsky.html.}} 
\citep{n09} also has imaging at the position of this field over the 
interval 2006--2008. No point
source is detected in a yearly sum of these images (see Table~\ref{tab:kait}). 

%\subsection{XMM/UV imaging}
%$XMM/UVOT has imaging in UVW1 and UVM1 in 2003. 
%$%It is will be a good reference for Swift/UVOT follow-up.

%\subsection{NVSS imaging}
%There is a 49.8\,mJy source roughly 50.6 arcsec away. 

\section{Analysis}
\subsection{SED}
We fit a blackbody spectrum to the optical and near-infrared fluxes
of PTF10\,fqs without taking into account any local extinction. 
The best fit gives a lower limit on the temperature of $\sim$3900\,K.

%\begin{figure}[htbp] 
%   \centering
%   \includegraphics[width=3.5in]{sed.ps} 
%   \caption{Spectral energy distribution (radio to X-ray) of PTF\,10fqs. 
%Upper limits are denoted by downward arrows. If we fit a blackbody
%without taking into account reddening, we get $T >\sim 3900$\,K and 
%$R \sim 3.8 \times 10^{14}$\,cm.}
%   \label{fig:sed}
%\end{figure}

\subsection{Spectral Modelling}

\begin{deluxetable}{llll}[!hbt]
  \tabletypesize{\footnotesize}
  \tablecaption{PTF\,10fqs Spectrum Properties}
  \tablecolumns{4}
  \tablewidth{0pc}
 \tablehead{\colhead{Line} & \colhead{Obs $\lambda$} &
             \colhead{Flux} & \colhead{Eq. Width} \\
\colhead{} & \colhead{(\,\AA\,)} & \colhead{(erg cm$^{-2}$ s$^{-1}$)} & \colhead{\,\AA\,} 
} 
  \startdata
H$\alpha$ & 6621.2 & $1.0 \times 10^{-15}$ & $-$19.9 \\ %26.5
H$\beta$  & 4907.3 & $1.3 \times 10^{-16}$ & $-$3.7  \\ %15.5
Na~I~D    & 5939.0 & $-3.1 \times 10^{-16}$ & 6.4 \\ %14.7
$[$Ca~II$]$   & 7355.8 & $2.9 \times 10^{-16}$ & $-$6.1 \\ %20.3
$[$Ca~II$]$   & 7387.2 & $1.8 \times 10^{-16}$ & $-$3.7 %19.8
  \enddata
  \tablecomments{Above line fluxes are measured on combined HET spectra (phase between +5\,days and +17\,days)}
\label{tab:lines}
\end{deluxetable}

We combined the four spectra obtained with HET (between +5\,days and
+17\,days). The most prominent (narrow) features in the spectra of
PTF\,10fqs are H$\alpha$, [Ca~II], the Ca~II near-IR triplet, Na~I~D, and
H$\beta$. The measured line fluxes and equivalent widths are summarized in
Table~\ref{tab:lines}. The H$\alpha$ FWHM is $\approx$930\,km s$^{-1}$ 
(taking into account the instrumental resolution). 

The Ca~II near-IR triplet is of particular interest. The HET spectra appear to
show a flux excess longward of 8300\,\AA\ beyond that expected from a
simple, low-order polynomial fit to the continuum. Together with a 
possible broad flux deficit near 8300\,\AA, the overall effect 
suggests a P-Cygni profile. If we fit three Gaussians, the Ca~II near-IR 
triplet features are broader than the [Ca~II] doublet,
%$\lambda\lambda$ 7291.47, 7323.89, 
and quite likely even broader than the narrow component of the H$\alpha$ 
profile. There is a surplus of flux at 8600\,\AA,
which falls right between the 8498.02, 8542.09\,\AA\ pair and the more isolated
8662.14\,\AA\ line, such as one would expect from an underlying broad feature. 

\begin{figure}[htbp] 
   \centering
   \includegraphics[width=3.5in]{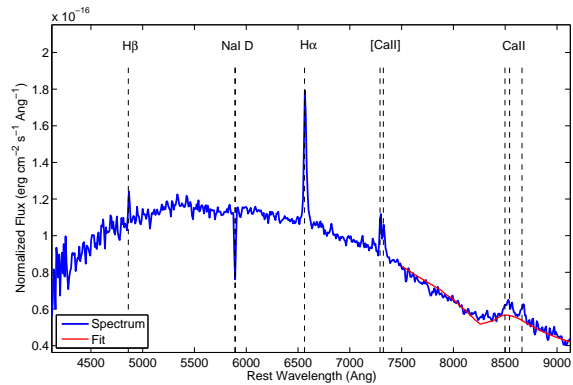} 
   \caption{SYNOW fit to summed HET spectra of PTF\,10fqs. Note the 
broad, possibly P Cygni, feature under the Ca~II near-IR triplet.}
   \label{fig:synow}
\end{figure}

We test this hypothesis further with SYNOW \citep{jb90} modelling.
We do not get a good fit to the overall shape of the spectrum with
an extinguished blackbody of any temperature (assuming standard dust).
To fit the red end of the spectrum, we need high temperature and
extinction (consistent with the strong Na~I~D absorption).
We find that in addition to narrow emission from Ca~II~IR, there
is also a likely underlying broad component (see Figure~\ref{fig:synow}).
The width (FWHM) of this feature is $\approx$10,000\,km s$^{-1}$.

A caveat to this interpretation is that a similar broad feature is not
seen in the H$\alpha$ profile. However, as noted below
(\S\ref{sec:TheSpectrum}), reinspection
of the spectra of related transients shows possible evidence of
a similar broad feature. Thus, we cautiously accept the interpretation 
that in addition to the low-velocity outflow seen in H$\alpha$, there 
is a higher velocity outflow in this explosion.

\section{What is PTF\,10fqs?}
In a nutshell, PTF\,10fqs is a red transient with a peak luminosity 
of $M_r = -$12.3 and a spectrum dominated by H$\alpha$, [Ca~II], and
Ca~II emission.  The width of the H$\alpha$ line is
$\approx$930\,km s$^{-1}$, and there is some evidence for a $\approx$10,000\,km
s$^{-1}$ broad Ca~II IR feature. %The light curve appears to be
%evolving slowly.% over the first three weeks.

The peak absolute magnitude and the H$\alpha$ line width of PTF\,10fqs are similar
to those seen in M85OT2006-1 (hereafter, M85-OT; \citealt{kor+07}), 
SN\,2008S \citep{pkt+08,sgc+09}, and NGC\,300-OT \citep{bbb+09, bsc+09}. 
However, there are some differences amongst these four
sources. %(and a closely related event, M31-RV; \citealt{rmp+89}). 
Thus, to aid a better classification, we review the similarities 
and differences between these four sources.

\subsection{The Light Curve}
The light curves of all four transients 
(PTF\,10fqs, SN\,2008S, NGC\,300-OT, and M85-OT) were red and 
evolved slowly for the first couple of months. %(e.g. \citealt{bps+09}). 
PTF\,10fqs had a well-sampled rise (Figure~\ref{fig:lc}) --- 
it rose by 1.1\,mag in r-band 
in 10.8\,days. After maximum, PTF\,10fqs declined slowly in r-band by 
1\,mag in 68\,days. Subsequently, it evolved more rapidly, declining by 
the next 1.3\,mag in 16\,days. PTF\,10fqs had $g-r$=1.0 at peak and declined 
relatively faster in $g$-band (1\,mag in 40\,days) than $r$-band.
In comparison, SN\,2008S declined by 1\,mag in 51\,days in $R$-band and
44\,days in $V$-band. The epoch of maximum light is uncertain for NGC\,300-OT 
due to lack of observations and is constrained to be anywhere between April 24 and 
May 15, 2008 \citep{bbb+09}. If we assume it to be April 27, 
the evolution is $R$-band and $I$-band are similar to that for 
PTF\,10fqs (Figure~\ref{fig:lc}). 

%The colors are red with $g-r > 1.5$ mag. PTF\,10fqs is also very red, 
%was caught on the rise and the light curve appears to be evolving slowly.

\begin{figure*}[htbp] 
   \centering
   \includegraphics[height=6in, angle=90]{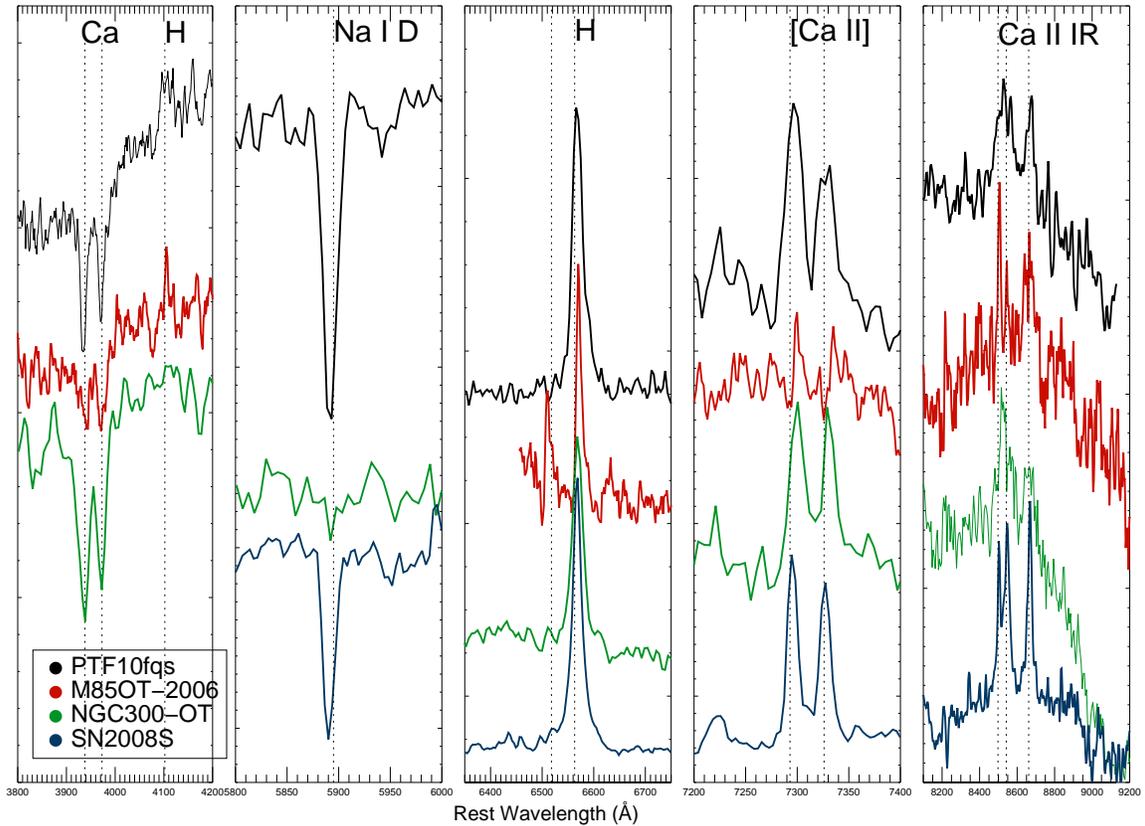} 
   \caption{Comparison of specific lines in spectra of PTF\,10fqs (black),
M85-OT (red; \citealt{kor+07}), SN2008s (blue; \citealt{bps+09}) and NGC\,300-OT (green; \citealt{bbb+09}). 
{\it From left to right:} Panel 1 shows Ca~II H\&K in
all three transients. Panel 2 shows the extreme Na I D absorption
in PTF\,10fqs. Panel 3 shows the similar H$\alpha$ widths in all three
transients. Note the presence of Fe~II in M85-OT. Panel 4 shows narrow 
[Ca~II] in all three transients. Panel 5 shows Ca~II near-IR triplet.
Note that in addition to the narrow lines, there is possibly an underlying
broad feature.}
   \label{fig:speczoom}
\end{figure*}

\subsection{The Spectrum}
\label{sec:TheSpectrum}

The spectral evolution of SN\,2008S \citep{bps+09} and NGC\,300-OT
\citep{bsc+09} were very well studied as they were in very nearby
galaxies. We took this opportunity to 
reanalyze the spectrum of M85-OT reported by
\citep{kor+07}{\footnote{In addition
to the features mentioned by \citealt{kor+07}, we can securely
identify Ca~II H\&K and see evidence of [Ca~II] and the Ca~II near-IR triplet.
Furthermore, we can identify the lines previously marked
``unidentified'': 4115\,\AA\ is H$\gamma$, 6428\,\AA\ is likely Fe~II
(multiplet 74), 6527\,\AA\ is likely Fe~II (multiplets 40 and 92).}}. 

Armed thus, we compare
and contrast the spectral features of these four transients (see
Figure~\ref{fig:speczoom}).

%1. The Mystery Lines of M85-OT
%+ 4115A @ z = 0.0024 => 4105A -- Isn't this just H-gamma (rest=4102A)?

%+ 6428A @ z = 0.0024 => 6413A -- Could this be Fe II (74) which is 6416.9A?
%This is in the dichroic gap of the Feb 24 spectrum plotted in figure, where
%do you measure it?

%+ 6527A @ z = 0.0024 => 6511A -- Any chance this is the Fe II (40) + Fe II
%(92) which are at 6516.1 + 6517.0? The slight blue-shift of the Fe II lines
%may be that the velocity I am using at this position is different
%from the galaxy or wavelength calibration?

%+ 8079A, 8106A. No idea. Are these good in the 2d spectrum? Could you
%point me to the 2d data please?

%2. Additional Lines: Ca H+K, [Ca II], Ca II IR:
%See attached zoom-in to the spectra, there
%is evidence for all three being present in the spectrum.

%3. Are you confident of the upper limit on H-beta flux of 6x10^-18 cgs
%on Feb 3? This would suggest H-beta flux increased by a factor of 15
%between Feb 3 and Feb 24... Not consistent with SN\,2008S/NGC\,300-OT.

\begin{itemize}

\item The H$\alpha$ profile of SN\,2008S showed a narrow component 
(unshocked circumstellar material [CSM]; $\approx$250\,km s$^{-1}$),
an intermediate component (shocked material between the ejecta and
the CSM; $\approx$1000\,km s$^{-1}$ ), and a broad component 
(underlying ejecta emission; $\approx$3000\,km s$^{-1}$).
NGC\,300-OT exhibited narrow (560\,km s$^{-1}$) and intermediate-width
components (1100\,km s$^{-1}$).  M85-OT only had a narrow component
(350\,km s$^{-1}$). PTF\,10fqs shows an intermediate-width component
(930\,km s$^{-1}$) in the H$\alpha$ emission line.

\item SN\,2008S had an H$\alpha$/H$\beta$ ratio that
evolved from 4 to 10. NGC\,300-OT had a ratio of 6, while M85-OT showed a
ratio of 3.5.  PTF\,10fqs has a ratio of 6.5. All events show flux
ratios higher than 3.1 (the expectation from Case B recombination). This may
be evidence for collisional excitation \citep{du80}.

\item PTF\,10fqs, NGC\,300-OT and SN\,2008S exhibit three calcium features: Ca~II H\&K 
in absorption, [Ca~II] and Ca~II near-IR triplet in emission. A reanalysis
of M85-OT shows Ca~II H\&K, as well as lower signal-to-noise ratio 
detections of both [Ca~II] and Ca~II IR. 
\citealt{sgc+09} show a similarity between the spectra of SN2008S and a 
Galactic hypergiant (IRC+10420) and suggest that strong [Ca~II] is due 
to destruction of dust grains.

%In 2008S, Ca II IR shows decreasing redshift with time. 
%In NGC\,300-OT, [Ca II] lines were significantly narrower than Ca II IR, 
%did not have a blue wing and decreased in width with time. 
%Thus, Ca II NIR and [Ca II] likely originate in physically distinct regions. 

%The presence of strong [Ca II] is consistent
%with narrow line widths of Balmer lines since if Ly$\alpha$ was broader than
%800 km s$^{-1}$, it would suppress ionization from Ca$^{+}$ to Ca$^{++}$.

\item As noted earlier (see also Figure~\ref{fig:synow}), there is
evidence for a broad feature around 8600\,\AA\ in the spectrum of
PTF\,10fqs. Motivated by this finding, we reinspected the spectra of
previous transients and find that a similar broad feature may also 
be present in the spectra of M85-OT and NGC\,300-OT.

\item Narrow Fe~II lines are visible in NGC\,300-OT and SN\,2008S.  
Reanalysis of M85-OT spectra possibly shows Fe~II(74) and Fe~II (40, 92).

\item For SN\,2008S, Na~I~D evolves from strong absorption at early times
to emission at very late times. This suggests a very dense CSM. 
O~I $\lambda$7774 is also in emission at late times.  For NGC\,300-OT, 
Na~I~D has a much lower equivalent width at early times, but it also 
evolves from absorption to emission. Neither Na~I~D nor O~I are seen in
M85-OT, but there is possibly K~I in emission. PTF\,10fqs has an
equivalent width of Na~I~D of 6.4, higher than SN\,2008S (2.3--4.4)
and NGC\,300-OT (1.0--2.1). The equivalent width of Na~I~D is too 
high to apply a standard correlation to estimate extinction.

\end{itemize}

\begin{figure}[htbp] 
   \centering
   \includegraphics[width=3.5in]{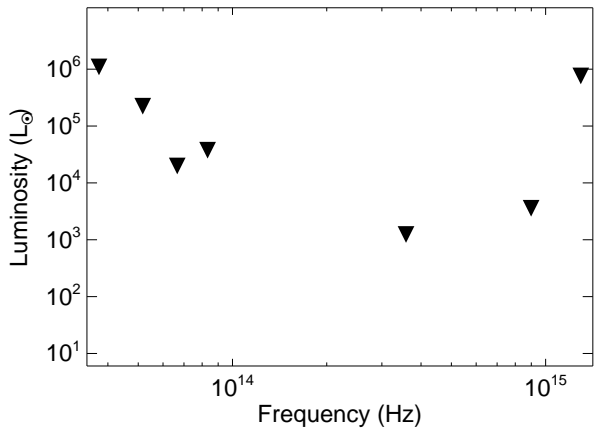} 
   \caption{Spectral energy distribution (mid-IR to UV) constraints 
on the pre-explosion counterpart of PTF\,10fqs. Upper limits are denoted 
by downward arrows.}
   \label{fig:sedprog}
\end{figure}

\begin{figure}[htbp] 
   \centering
   \includegraphics[width=3.5in]{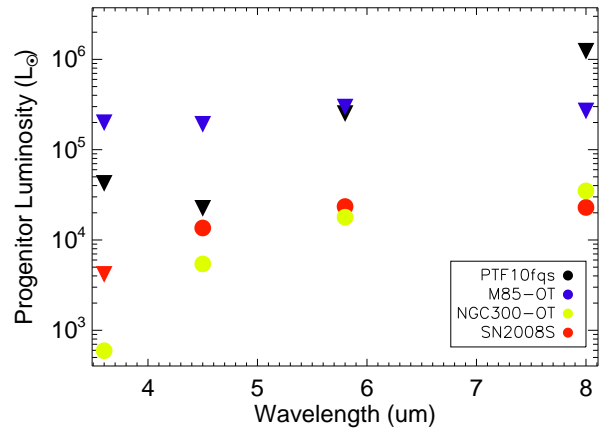} 
   \caption{Pre-explosion detections (circles) or upper limits 
(downward triangles) from {\it Spitzer} for PTF\,10fqs, SN\,2008S, 
NGC\,300-OT, and M85-OT. The non-detection of a progenitor for PTF\,10fqs
and M85-OT does not rule out a progenitor of luminosity comparable to that
detected for NGC\,300-OT and SN\,2008S.}
   \label{fig:prog_mir}
\end{figure}

\subsection{The Pre-Explosion Counterpart}
We plot the upper limits on the pre-explosion counterpart for PTF10fqs
in Figure~\ref{fig:sedprog}. The most constraining limits are in the 
optical. %At the distance of M99, the {\it HST}/WFPC2 limits for PTF\,10fqs rule 
%out O/B progenitors fainter than $M_V = -4.3$\,mag and K/M supergiants 
%fainter than $M_V = -3$\,mag (colors from \citealt{c00}). There may be 
%some extinction at the site of PTF\,10fqs. Even with $A_V$ of up to 1\,mag, 
%we can rule out red supergiants, blue supergiants, and main-sequence 
%stars earlier than B2. 
Following the Geneva stellar evolution tracks \citep{ls01} for unenshrouded stars, 
the luminosity limit of M$_I > -$4.3 corresponds to a progenitor mass $<$4\,M$_{\odot}$.
If there was extinction of, say 1.5\,mag, this would change the limit to $<$7\,M$_{\odot}$.
None of SN\,2008S, NGC\,300-OT, M85-OT, and PTF\,10fqs have an optical
counterpart in deep, pre-explosion optical images.
The limits in all cases are deep enough to at least rule out red 
supergiants and blue supergiants.

For both SN\,2008S and NGC\,300-OT, an extremely red and luminous mid-infrared
pre-explosion counterpart is seen \citep{pkt+08,tps+09}.
Recently, \citet{ksp+10} show that such progenitors
are as rare as 1 per galaxy (and possibly associated
with a very short-lived phase of many massive stars).
Thus, both of these transients can be reasonably associated
with massive stars. 
Unfortunately, the large distance to M85 and M99 means that the
pre-explosion {\it Spitzer} limits on M85-OT and PTF\,10fqs are not
deep enough by a factor of few to constrain their progenitors to
similar depths (see Figure~\ref{fig:prog_mir}).  %(M31-RV exploded
%before {\it Spitzer} was launched.)

%The 4.5\,$\mu$m absolute magnitude of the progenitors 
%of NGC\,300-OT was $-$10.5 and SN\,2008S was $-$11.5. Unfortunately, the limit
%of $-$14.2 is not deep enough to test whether PTF\,10fqs had a similar MIR 
%progenitor.

%\begin{figure}[htbp] 
%   \centering
%   \includegraphics[width=3in, angle=0]{SNF-1.eps} 
%   \caption{Multiplicity of supernovae in galaxies. 
%This histogram is based on
%the compilation of supernovae by the 
%IAU (\texttt{http://www.cfa.harvard.edu/iau/lists/Supernovae.html}).
%Note that this sample includes supernovae discovered in the past century
%from many different sources and the inherent selection biases have not been
%taken into account.}
%   \label{fig:firy}
%\end{figure}

\subsection{The Large-Scale Environment}

M85-OT is located in the lenticular galaxy M85 (also in the
Virgo cluster). Fortunately, this galaxy was observed with {\it HST}
for the ACS Virgo Cluster Survey as well as for a GO program.
The transient is not associated with any star-forming region and
the absolute magnitude of the progenitor is fainter than 
$M_{g} \approx -4$ ($<$7\,M$_{\odot}$ not correcting for extinction; 
\citealt{okr+08}). Thus, a massive-star origin is quite unlikely.

In contrast, SN\,2008S, NGC\,300-OT, and PTF\,10fqs occurred in 
star-forming galaxies. It may be worth noting here that three supernovae 
(all of the core-collapse variety) have previously been discovered in 
the host galaxy of PTF\,10fqs. %In Figure~\ref{fig:firy}, we show a
%histogram of the multiplicity of known supernovae. 
It is perhaps of some significance that eight supernovae (six core-collapse, two
unclassified) were discovered in NGC\,6946 in addition to SN\,2008S.
Only one supernova (of Type Ia) has been discovered in NGC\,300.
Small-number statistics and discovery bias (incompleteness from variety of 
different searches) notwithstanding, we make the suggestion that galaxies with 
a high supernova rate preferentially produce luminous red novae. If this suggestion 
is correct, then it would be worth the effort to systematically maintain close vigilance 
on the nearest galaxies having large supernova rates.

\citealt{kor+07} suggested that V838\,Mon, V4332~Sgr and M31~RV may also
be luminous, red novae. We note here that the two Galactic sources are 
located in star-forming regions. Specifically, V838\,Mon
is in a young (25\,Myr) star cluster and may even
have a B3 companion \citep{ab07}. V4332\,Sgr \citep{mwt+99} is located towards
the inner Galaxy (in Sagittarius).  On the other hand, M31~RV is located in the 
bulge of M31. {\it HST} observations (undertaken
with WFPC2 in parallel mode) taken about a decade ago show that the
immediate environs of M31-RV are  typical bulge-population stars
\citep{bs06}. No unusual remnant star is seen at the astrometric
position of M31~RV, nor any evidence of a light echo (consistent with
the absence of dense circumstellar or interstellar gas that is 
essential to form echoes). Separately, there is no evidence for any 
luminous outbursts in this area in the period 1942--1993 \citep{bm04}.  
Thus, M31~RV appears to have been a cataclysmic event in the bulge of M31.

%\Subsection{Energetics}
%The energetics for eAGB explosions and LRNe are similar --- 
%10$^{47}$ ergs for NGC\,300-OT and 6$\times$10$^{46}$ ergs for M85-OT.
%Need bolometric light  curve to compute for PTF\,10fqs

%\section{Summing up}
%To first order, the explosion signature (luminosity, light curve, spectra) 
%of PTF\,10fqs is similar to the comparison set of M85-OT,
%NGC\,300-OT, and SN\,2008S. The principal difference arises in
%the pre-explosion counterpart (the progenitor) and the large-scale
%environment. 
%
%Specifically, NGC\,300-OT and SN\,2008S are remarkable for their
%very bright progenitors. %There were no deep pre-explosion
%%observations of M31\,RV. 
%Though sensitive pre-explosion observations
%of M85-OT and PTF\,10fqs do exist, the large distance to the Virgo Cluster 
%(17\,Mpc) relative to that of NGC\,300 (1.9\,Mpc) and NGC6946 (5.7\,Mpc) 
%results in weak constraints on the luminosity of any pre-explosion star.
%
%Thus, we conclude that PTF\,10fqs is, at the very least, a rare
%transient like M85-OT, NGC\,300-OT, and SN\,2008S. The star-forming
%location of PTF\,10fqs makes it more similar to NGC\,300-OT and SN\,2008S.

\section{Conclusion}

PTF\,10fqs is the fourth member of a class of extragalactic transients\footnote{Henceforth
we use the term ``luminous red novae'' as a functional short name for
such events.} which possess a peak luminosity between that of novae and 
supernovae, and have spectral and photometric evolution that bear no resemblance
to either supernovae or novae. The other members of this class are M85-OT, NGC\,300-OT,
SN\,2008S. 

NGC\,300-OT and SN\,2008S are remarkable for their very bright mid-infrared 
progenitors. Though sensitive pre-explosion observations
of M85-OT and PTF\,10fqs do exist, the large distance to the Virgo Cluster 
(17\,Mpc) relative to that of NGC\,300 (1.9\,Mpc) and NGC6946 (5.7\,Mpc) 
results in weak constraints on the luminosity of any pre-explosion star.
PTF\,10fqs, NGC\,300-OT, and SN\,2008S occurred in star-forming regions whereas 
M85-OT was in the bulge. {\it Prima facie}, this group of explosive events can be 
divided into a disk and a bulge group.

The discovery of PTF\,10fqs in itself cannot address whether the two
groups of luminous, red novae are one and the same. 
The proposed models to explain this group are diverse: electron
capture within an extreme asymptotic giant branch (AGB) star, 
common-envelope phase (stellar merger), inspiral of a giant planet 
into the envelope of an aging parent star, a most peculiar nova, 
and a most peculiar supernova.

The possible evidence of the broad feature centered around the Ca~II 
near-IR triplet with an inferred velocity spread of 10,000\,km s$^{-1}$ 
may be an important clue. It would mean
that these events possess both a low- and a high-velocity outflow. By
comparison with other astronomical sources, one can envisage a high-velocity
polar outflow and a slower equatorial outflow (but with a larger
mass).  To this end, continued sensitive spectroscopy of PTF\,10fqs
(and of course other such future events) would be very valuable. 

The ``Transients in the Local Universe'' key project of the Palomar
Transient Factory is designed to systematically unveil events in
the gap between novae and supernovae. It surveys $\approx$20,000 nearby
galaxies ($d < 200$\,Mpc) yearly at 1-day cadence and a depth of
$R < 21$ mag.  (If the maximum luminosity of this class is 
$-14$ mag, then we would be sensitive to events out to 100\,Mpc.) Furthermore, 
{\it Spitzer} has a growing archive of deep images of nearby galaxies (e.g., 
SINGS, \citealt{kab+03}; LVL, \citealt{dcj+09}, and 
S4G, \citealt{srh+10}), and 
{\it WISE} \citep{wem+10} has an ongoing all-sky survey in the mid-IR. This will 
allow us to probe deeper in search of the pre-explosion counterpart and possibly
present a new channel for discovery of luminous red novae.  The discovery of 
PTF\,10fqs is only the harbinger of the uncovering of a large sample of such transients
to unveil the nature of this new class of explosions.

%Good idea: TILU. Write a little summary of what the TILU key project is and then e
%nd the paper with TILU is working!

%The simplest answer lies in the discovery of another member of this class 
%that not only shares the explosion signature, but also has a pre-explosion
%eAGB progenitor and is located in the bulge of its host.

\bigskip
\smallskip

\noindent{\bf Acknowledgments.} 

M.M.K. thanks the Gordon and Betty Moore Foundation 
for a Hale Fellowship in support of graduate study. 
The Weizmann Institute PTF participation is supported in part by
the Israel Science Foundation via grants to AGY. The Weizmann-Caltech
collaborative PTF effort is supported by the US-Israel Binational Science 
Foundation. AGY and MS are jointly supported by the ``making connections'' 
Weizmann-UK program. AGY further acknowledges support by a Marie Curie IRG 
fellowship and the Peter and Patricia Gruber Award, as well as funding
by the Benoziyo Center for Astrophysics and the Yeda-Sela center at the
Weizmann Institute. A.V.F.'s group and KAIT are supported by National Science Foundation
(NSF) grant AST-0908886, the Sylvia \& Jim Katzman Foundation, the Richard \& Rhoda Goldman Fund, 
Gary and Cynthia Bengier, and the TABASGO Foundation;
additional funding was provided by NASA through {\it Spitzer} grant
1322321, as well as {\it HST} grant AR-11248 from
the Space Telescope Science Institute, which is operated by Associated
Universities for Research in Astronomy, Inc., under NASA contract NAS
5-26555. J.S.B. and his group are partially funded by a DOE
SciDAC grant. E.O.O. and D.P. are supported by the Einstein fellowship.
L.B. is supported by the National Science Foundation under grants PHY 05-51164 and
AST 07-07633.

We are grateful to the staff of the Gemini Observatory for their
promptness and high efficiency in attending to our TOO request.
Likewise, we thank the staff of the Very Large Array and
the Hobby Eberly Telescope. We acknowledge the following internet
repositories: SEDS (Messier Objects) and GOLDMine (Virgo Cluster),
Finally, as always, we are grateful to the librarians who maintain
the ADS, the NED, and SIMBAD data systems.

The Hobby-Eberly Telescope (HET) is a joint project of the University
of Texas at Austin, the Pennsylvania State University, Stanford
University, Ludwig-Maximillians-Universit\"at M\"unchen, and
Georg-August-Universit\"at G\"ottingen. The HET is named in honor
of its principal benefactors, William P. Hobby and Robert E. Eberly.
The Marcario Low-Resolution Spectrograph is named for Mike Marcario
of High Lonesome Optics, who fabricated several optics for the
instrument but died before its completion; it is a joint project
of the Hobby-Eberly Telescope partnership and the Instituto de
Astronom\'{\i}a de la Universidad Nacional Aut\'onoma de M\'exico.
{\it GALEX} (Galaxy Evolution Explorer) is a NASA Small Explorer, 
launched in 2003 April. We gratefully acknowledge NASA's support for
construction, operation, and science analysis for the {\it GALEX} mission,
developed in cooperation with the Centre National d'Etudes Spatiales
of France and the Korean Ministry of Science and Technology. PAIRITEL is operated 
by the Smithsonian Astrophysical Observatory (SAO) and was made possible
by a grant from the Harvard University Milton Fund, the camera loan from the 
University of Virginia, and the continued support of the SAO and UC Berkeley.
The Expanded Very Large Array is operated by the National Radio Astronomy
Observatory, a facility of the NSF operated
under cooperative agreement by Associated Universities, Inc.

\bigskip

%\bibliographystyle{apj}
%\bibliography{M99}

\end{document}